# Coherent phononic frequency combs in ferroelectric CMOS oxides

Jinghan Gao[1,a], Shruti Mishra[1,a], Yilin Kou[2],
S M Enamul Hoque Yousuf[1], Hanna Cho[2], Roozbeh Tabrizian[1*]

[1]*Department of Electrical Engineering and Computer Science, University of Michigan, Ann Arbor, 48109, MI, USA*

[2]*Department of Mechanical and Aerospace Engineering, Ohio State University, Columbus, 43210, OH, USA*

**To whom correspondence should be sent: rtabrizi@umich.edu*

*a. These authors contributed equally: Jinghan Gao, Shruti Mishra.*

## Abstract

Modern electronic systems require tens of clock and carrier frequencies, each synthesized by a dedicated phase-locked loop from a shared reference, imposing routing, power and synchronization burdens that grow with every domain. Optical frequency combs solved this problem in photonics, whereas electronics has lacked an equivalent source in its native radiofrequency domain. Here we report broadband phononic frequency combs in ferroelectric hafnia–zirconia nanoelectromechanical resonators built from complementary metal–oxide–semiconductor (CMOS) oxides. Lithographically defined detuning of a 2:1 internal resonance selects the generation mechanism: two-tone-seeded wave mixing yields more than 170 lines distributed over two octaves, with mutual coherence verified for representative pump and generated lines, whereas an autonomous Hopf route yields hierarchical combs of more than 200 lines through torus and period-doubling dynamics, in agreement with slow-flow bifurcation theory. Geometric scaling extends comb generation across 0.44–2.1 GHz. Heterodyne measurements, analyzed using the modified Allan deviation (MDEV), establish a two-timescale law. The mechanism governs short-term stability: seeded combs inherit the white-phase-noise scaling of their pumps, whereas autonomous combs acquire the phase diffusion of a free-running oscillator, with the one-second MDEV increasing from $10^{-11}$ to $10^{-8}$. The material governs long-term stability: in air and without active thermal control, the temperature-compensated stack suppresses the random-walk drift that dominates uncompensated resonators. These results establish mechanism- and material-level design rules for operating a single resonator as chip-scale frequency infrastructure, from multi-clock generation to radiofrequency parallel processing.

## Introduction

Integrated electronic systems are increasingly defined by frequency multiplicity rather than by a single master clock. Radiofrequency transceivers, data converters, digital processing islands, memory interfaces and sensor front ends each operate in a clock domain of their own[1–3], where a contemporary system-on-chip contains tens of such domains, and chiplet-based integration extends timing distribution across dies[4]. This demand is universally served by a single stable reference, typically a quartz or microelectromechanical oscillator [5,6], whose output is routed across the system and multiplied to each target frequency by a dedicated phase-locked loop (PLL). The architecture carries a fundamental scaling burden. Reference distribution consumes routing, buffering and power while accumulating skew. Frequency multiplication by a factor $N$ amplifies the reference phase noise by $20\log N$ within the loop bandwidth[7]. Moreover, because every domain is synthesized through an independent loop, synchronization circuitry is required to recover the mutual coherence needed for interleaved data conversion, phased arrays, and die-to-die links, and its complexity grows with each additional domain.[8]. What these systems demand of a frequency source, moreover, is not arbitrary synthesis, which PLLs already provide: it is a dense arithmetic grid of mutually coherent references, $f_n = f_p + nf_r$ , of the kind consumed by beamforming, interleaved conversion, multi-carrier transceivers, sensor-array readout and chiplet synchronization, delivered from one integrated device. The difficulty is therefore no longer generating one stable frequency, which mechanical resonators address well; it is providing many mutually coherent frequencies[9] from a single integrated source.

Optical frequency combs established the template for such a source. A driven nonlinear cavity can generate hundreds of equidistant spectral lines whose mutual coherence, defined as the deterministic rather than diffusive evolution of their relative phases such that any pairwise beat note remains spectrally sharp, arises from nonlinear interaction rather than by external locking[10–12]. Optical frequency comb technology revolutionized precision frequency metrology, an advance recognized by the 2005 Nobel Prize in Physics, and combs today serve not only as clocks, synthesizers and spectral rulers[13–15] but as parallel coherent carriers for dual-comb spectroscopy, ranging and photonic computing[16–18]. Optical combs, however, operate far from the frequencies and material stacks of electronic clocking and signal processing: their carriers lie at hundreds of terahertz, and every radiofrequency function they enable must pass through lasers, modulators and photodetection, carrying the integration and power overheads of microwave photonics[19]. A direct analog for electronics is the phononic frequency comb (PFC), in which nonlinear coupling between the acoustic modes of a mechanical resonator generates equidistant lines natively in the kilohertz–gigahertz range in the electrical domain through integrated electromechanical transduction[20,21].

PFCs have been demonstrated over the past decade through intrinsic three-wave mixing[20], parametric pumping[22], bifurcations of driven modes[21,23], mechanical overtone dynamics[24], soliton-like states[25], multimode interactions[26,27] in piezoelectric micromachined transducers[28], quartz bulk-acoustic-wave devices[29,30] and optomechanical self-oscillation[31,32]. Yet the field has not followed the trajectory of its optical counterpart, for reasons that are instructive. Most demonstrations rely on electrostatically or piezoelectrically transduced silicon microstructures, quartz crystals, dielectric membranes, or optically transduced cavities, which do not offer a route to monolithic integration with CMOS electronics. Weak intrinsic nonlinearities confine most combs to narrow bands around a single mode, with line counts of a few tens, unless engineered mode-coupling structures or elevated drive powers are employed. Comb existence occupies narrow regions of drive frequency and amplitude corresponding to subsets of the Arnold tongue of the underlying internal resonance (InRes)[33], and these regions have largely been located empirically rather than by design. Most importantly, while comb spectra are routinely reported, the frequency stability of the comb lines, which ultimately determines their utility, has been examined only in isolated cases, such as a resonance-tracking tool for sensing[34] or as a single figure of merit[31,35], and the connection between PFC generation mechanisms and the stability of their lines has not been systematically established across distinct nonlinear states. In optics, this connection is central: modulation-instability comb states and dissipative Kerr soliton states can produce broadly similar optical spectra while exhibiting fundamentally different coherence[15]. Identifying and controlling the low-noise states is the key step in transforming microresonator combs into technology, and noise-transfer theory now links the nonlinear state quantitatively to timing jitter[36] and system-level frequency synthesis[13]. For PFC, the corresponding chain remains incomplete: stochastic analyses of the noisy Hopf normal form determine when a limit cycle remains phase-coherent[37], but few quantitative models or experimental studies have connected the PFC's generation mechanism to the stability of its lines. Establishing this connection would transform PFCs from demonstrations of nonlinear dynamics into engineerable sources of coherent frequency grids for electronic systems.

Ferroelectric hafnium–zirconium oxide ($Hf_{0.5}Zr_{0.5}O_2$; HZO) provides a platform to close this gap. HZO is grown by atomic layer deposition within CMOS-compatible thermal budgets and has been developed from ultrathin electromechanical transducers[38] to gigahertz HZO-alumina ($Al_2O_3$) superlattice resonators[39], monolithic switchable filter arrays[40] and, most recently, temperature-insensitive mechanical resonators and clocks in which the phase-transformation-induced elastic anomaly of HZO is combined with silicon dioxide ($SiO_2$) to compensate the first- and second-order temperature coefficients of frequency simultaneously[41]. Those advances established HZO–$Al_2O_3$–$SiO_2$ as a CMOS-native stack for linear frequency control. This article exploits the opposite face of the same material system: the strong intrinsic nonlinearity that degrades its linear devices[38] but constitutes the essential resource for parametric frequency generation. Four properties make the platform uniquely suited to comb generation. The ferroelectric and electrostrictive responses supply quadratic and cubic electromechanical nonlinearities substantially stronger than those of conventional piezoelectrics such as aluminum nitride; and the ultrathin released beams contribute a second, geometric source of quadratic nonlinearity through their static curvature, which breaks the out-of-plane symmetry of the restoring force. The dense modal spectrum of these beams makes 2:1 InRes pairs abundant rather than exceptional, arising naturally between width-extensional and flexural modes, and between flexural modes themselves, without the H- or T-shaped coupling structures engineered in conventional microelectromechanical resonators[42–45]. Atomic-layer control of the stack and lithographic control of the in-plane dimensions then select among this abundance: the InRes condition of a designated mode pair — including its residual detuning — is programmed by design rather than found by search, from tens of megahertz to gigahertz frequencies. The temperature-compensating HZO–$Al_2O_3$–$SiO_2$ stack[41] also suppresses the environmental fluctuations that dominate the long-term stability of uncompensated mechanical references.

In this article, we show that HZO–$Al_2O_3$–$SiO_2$ nanoelectromechanical resonators generate broadband, dense and mutually phase-coherent PFCs, and establish InRes mismatch between two mode frequencies of $\omega_1$ and $\omega_2$, i.e., $\sigma_i = \omega_2 - 2\omega_1$, as a lithographically programmable selector of the comb-generation pathway. For a nearly commensurate 2:1 modal pair, with $\sigma_i \approx 0$, between a width-extensional mode at $\omega_2$ and a flexural mode at $\omega_1$, quadratic modal coupling produces a stable and strong energy-exchange state with a characteristic M-shaped frequency response. Under two-tone excitation, this internally resonant state mediates cascaded three-wave interactions, forming an effective four-wave-mixing ladder that produces more than 170 lines distributed over two octaves (35–153 MHz), with mutual coherence verified for representative pump and generated lines, with a repetition frequency (i.e., comb spacing) set deterministically — and hence programmed electronically — by the pump-tone separation. By contrast, for a 2:1 pair of flexural modes with a finite mismatch ($\sigma_i > 0$), one-tone excitation provides an autonomous route to hierarchical comb generation through successive bifurcations. The pump-locked periodic response first undergoes a Hopf bifurcation in the slow-flow dynamics, yielding a stable slow-flow limit cycle that modulates the pumped response into an autonomous frequency comb; a subsequent secondary Hopf (torus) bifurcation adds a second modulating frequency, producing hierarchical combs of more than 200 lines that ultimately terminate in chaos. Slow-flow analysis shows that finite mismatch, together with quadratic intermodal coupling, is sufficient to open the Hopf-instability window in the reduced-order model; experimentally, lithographic geometry selects both the modal configuration and its residual mismatch. Combined with numerical continuation and Floquet stability analysis, the theoretical model further reveals the pathways to comb generation and the parameters governing the resulting comb characteristics (**Supplementary Information 1**). These results transform PFC generation from empirical tuning into a geometry- and drive-programmable design space. Scaling the InRes condition to a 15-nm stack and one-fifth in-plane dimensions extends comb operation to a 1.07 GHz width-extensional mode, with comb clusters distributed from approximately 0.44 GHz to 2.1 GHz.

The stability of a PFC, we find, is set by its nonlinear origin. By heterodyne down-conversion against a stable reference, we track the frequency stability of individual comb lines across the full ladder of nonlinear states — the direct pump response, internally resonant subharmonic response, two-tone-seeded wave-mixing comb, and Hopf-generated autonomous comb — and the organizing quantity is the number of free phase parameters each state carries. A seeded comb carries none: every line frequency is an arithmetic combination of externally supplied tones, so its stability inherits from the pump source with SNR and mixing order influencing its value. An autonomous comb carries one: its spacing is an emergent, Hopf-generated limit-cycle frequency, so the comb remains internally coherent, while the spacing is not locked to any external reference. Hierarchical states add further phase coordinates, whereas period doubling changes the periodicity without introducing another continuous free phase. This phase ledger predicts, and our measurements confirm, the stability class of each observed dynamical state. Direct mutual-coherence measurements independently validate its premise. Representative pump and generated-line pairs were simultaneously down-converted and phase-tracked over 300 s; after removal of their deterministic linear phase evolution, the relative phase remained bounded and non-diffusive, while the phase-derived spacing residual showed no systematic drift. These measurements confirm mutual coherence and a common repetition rate among the tested lines. Every externally referenced state preserves the $\tau^{-3/2}$ white-phase-noise scaling of its pump, differing only by a signal-to-noise penalty, so that the seeded comb operates, in effect, as a passive mechanical frequency synthesizer. The autonomous Hopf comb, by contrast, retains the pump-referenced carrier but acquires a free-running spacing phase, resulting in the phase diffusion characteristic of an autonomous oscillator and increase in the one-second modified Allan deviation (MDEV) by approximately three orders of magnitude, from $10^{-11}$ to $10^{-8}$. This excess instability is consistent with a stochastic Hopf normal-form in which non-isochronicity converts amplitude fluctuations into phase fluctuations along the limit cycle.[37] Stability moreover acts as a diagnostic of the nonlinear pathway itself: the two line families of the hierarchical comb separate into distinct noise branches; the scaling of instability with comb order distinguishes intrinsic nonlinear noise, which accumulates linearly with order, from measurement limits; and two spectra with identical halved spacing, one reached by half-index mixing of the seeded comb, one by period doubling of the autonomous torus, occupy opposite stability classes because the former inherits the stability from pump source while the latter develops free-running phase dynamics of the autonomous torus.

At long averaging times the material platform governs. Measured in air and without active thermal control, comb lines generated in the temperature-compensated HZO–$Al_2O_3$–$SiO_2$ stack[41] show no random-walk upturn over the full measured averaging-time range, settling instead on a flicker-frequency floor. Under comparable excitation, the comb state in an uncompensated HZO resonator develops a random-walk frequency-noise signature within a few hundred seconds, consistent with ambient thermal fluctuations. This direct comparison identifies the compensation layers as the origin of the observed suppression. The same random-walk degradation appears in an uncompensated aluminum scandium nitride resonator[46,47], confirming that the effect is set by temperature sensitivity rather than by material choice. The oxide stack therefore passively suppresses temperature-driven long-term drift across the measured comb lines, whereas conventional temperature-compensated oscillators generally protect a single frequency through sensing and correction circuitry.

In PFCs, as in optical ones, the spectrum alone is therefore not sufficient: the generation mechanism and the material platform together determine whether a comb can serve as frequency infrastructure, and the platform presented here provides design control over both. A single HZO resonator driven by one or two tones thereby becomes a monolithic source of mutually coherent references to which the PLLs of a heterogeneous system can lock with strongly reduced multiplication ratios — relaxing the routed reference tree and per-domain synthesis burden of multi-clock systems while remaining entirely within the CMOS oxide toolbox. Because this grid is mutually coherent, electronically programmable and native to the electrical domain, the same resource extends beyond clocking — toward dual-comb radiofrequency channelization and correlation, comb-referenced sensing and frequency-multiplexed analog computing[16,17] — functions that optical combs reach only through electro-optic conversion.

## Geometry-programmed comb pathways

The resonators studied here are doubly clamped beams fabricated from CMOS oxides and metal electrodes: an atomic-layer-deposited HZO–$Al_2O_3$ superlattice transducer on a $SiO_2$ temperature-compensation layer (**Methods**) — referred to hereafter as HZO CMOS-oxide resonators — with resonance frequencies set jointly by the ALD-controlled stack and the lithographically defined in-plane dimensions.[39,41] Two features of the platform organize everything that follows. First, the released ultrathin beams develop a static out-of-plane curvature that breaks the symmetry of the restoring force and, together with the electrostrictive response of the ferroelectric layer, supplies an unusually strong quadratic nonlinearity at low drive.[38] Second, this nonlinearity acts on a dense modal spectrum, making 2:1 mode pairs abundant rather than exceptional: near-commensurate pairs arise naturally between width-extensional and flexural modes, and between flexural modes themselves, without the deliberately engineered H- or T-shaped coupling structures typically required in conventional microelectromechanical resonators.[43,48,49] This rich InRes landscape turns the platform into a systematic setting for asking how the configuration of an InRes pair shapes the nonlinear dynamics it can host. Abundance, however, is only half of the design space; the other half is selection. Because the flexural spectrum shifts with the in-plane dimensions and stack while the width-extensional frequency is fixed by the width, lithography programs which pair is commensurate and — decisively — the residual detuning $\sigma_i = \omega_2 - 2\omega_1$ of that pair.

Among the many available pairs, we select two that bracket the detuning axis (**Fig. 1**). In a high-aspect-ratio beam (width 38 μm, length 200 μm), the fundamental width-extensional mode at $f_2 = \omega_2/2\pi = 76.8$ MHz is nearly commensurate with a higher-order flexural mode at $f_1 = \omega_1/2\pi = 38.4$ MHz, yielding $\sigma_i \approx 0$. In a low-aspect-ratio beam (width 42 μm, length 150 μm), two out-of-plane flexural modes at $f_2$ =66.6 MHz and $f_1$ = 32.4 MHz form a near 2:1 pair with a finite mismatch ($\sigma_i > 0$). Although both pairs satisfy the approximate 2:1 InRes condition, with the higher-frequency mode externally driven, they evolve through fundamentally different nonlinear dynamics. For the nearly commensurate pair ($\sigma_i \approx 0$), energy transfer from the directly driven width-extensional mode to its flexural mode produces a characteristic M-shaped saturation response that is nearly symmetric about $f_2$ (**Fig. 1a**). By contrast, introducing a finite frequency mismatch ($\sigma_i > 0$) distorts the M-shape response into a single hardening-like resonance (**Fig. 1b**), qualitatively consistent with previous theoretical predictions.[50]

Guided by these experimentally observed response signatures, we describe the dominant dynamics using a reduced-order model (ROM) incorporating quadratic 2:1 intermodal coupling and weaker cubic nonlinearities:

$$\ddot{u}_1 + 2\mu_1\dot{u}_1 + \omega_1^2 u_1 + \alpha u_1 u_2 + \chi_1 u_1 u_2^2 = 0$$

$$\ddot{u}_2 + 2\mu_2\dot{u}_2 + \omega_2^2 u_2 + \beta u_1^2 + \chi_3 u_2 u_1^2 + \kappa_2 u_2^3 = F\cos\omega_d t$$

where $u_1$ and $u_2$ denote the lower- and higher-frequency modal coordinates; $\omega_j = 2\pi f_j$ and $\mu_j$ are their mode frequencies and damping coefficients; $\alpha$ and $\beta$ describe quadratic intermodal coupling; $\chi_1$ and $\chi_3$ describe cubic cross-modal coupling; $\kappa_2$ is the cubic self-nonlinearity of the driven mode; and $F$ and $\omega_d$ are the amplitude and angular frequency of the external drive. For the nearly commensurate pair ($\sigma_i \approx 0$), both experiments and slow-flow analysis show that the internally resonant response remains dynamically stable throughout the resonance bandwidth, with no Hopf bifurcation predicted or observed (**Supplementary Information 1 S1-3**). This behavior contrasts with that reported for the complementary 1:2 InRes configuration, in which direct excitation of the lower-frequency mode can produce a Hopf bifurcation near the center of the M-shaped response.[48,51,52] Here we show that finite internal-frequency mismatch enables a Hopf instability in the 2:1 configuration.

To identify the Hopf-instability region, we derive the averaged slow-flow equations and calculate the pump-locked fixed points as functions of the external pump detuning $\sigma_e = \omega_d - \omega_2$, internal frequency mismatch $\sigma_i$, and drive amplitude $F$. We then linearize the slow flow about each fixed point and evaluate the third Hurwitz determinant, $\Delta_3 = p_1 p_2 p_3 - p_3^2 - p_1^2 p_4$, where $p_1$–$p_4$ are the coefficients of the Jacobian characteristic polynomial $P(\lambda) = \lambda^4 + p_1\lambda^3 + p_2\lambda^2 + p_3\lambda + p_4$. Provided the remaining Routh–Hurwitz conditions are satisfied, $\Delta_3 > 0$ indicates a stable pump-locked state, $\Delta_3 = 0$ marks the Hopf boundary, and $\Delta_3 < 0$ identifies the oscillatory instability. Near exact commensurability, $\Delta_3$ remains positive throughout the resonance bandwidth. As $\sigma_i$ increases, however, $\Delta_3$ crosses zero at two values of $\sigma_e$, defining the onset of a finite Hopf-instability window. Mapping these crossings over drive amplitude reveals a minimum drive threshold and a widening detuning interval at higher excitation, in qualitative agreement with the measured existence range of the autonomous comb. The persistence of this boundary after removing the cubic terms demonstrates that finite mismatch $\sigma_i$ and quadratic intermodal coupling are sufficient to produce the Hopf instability (**Supplementary Information 1 S1-3**).

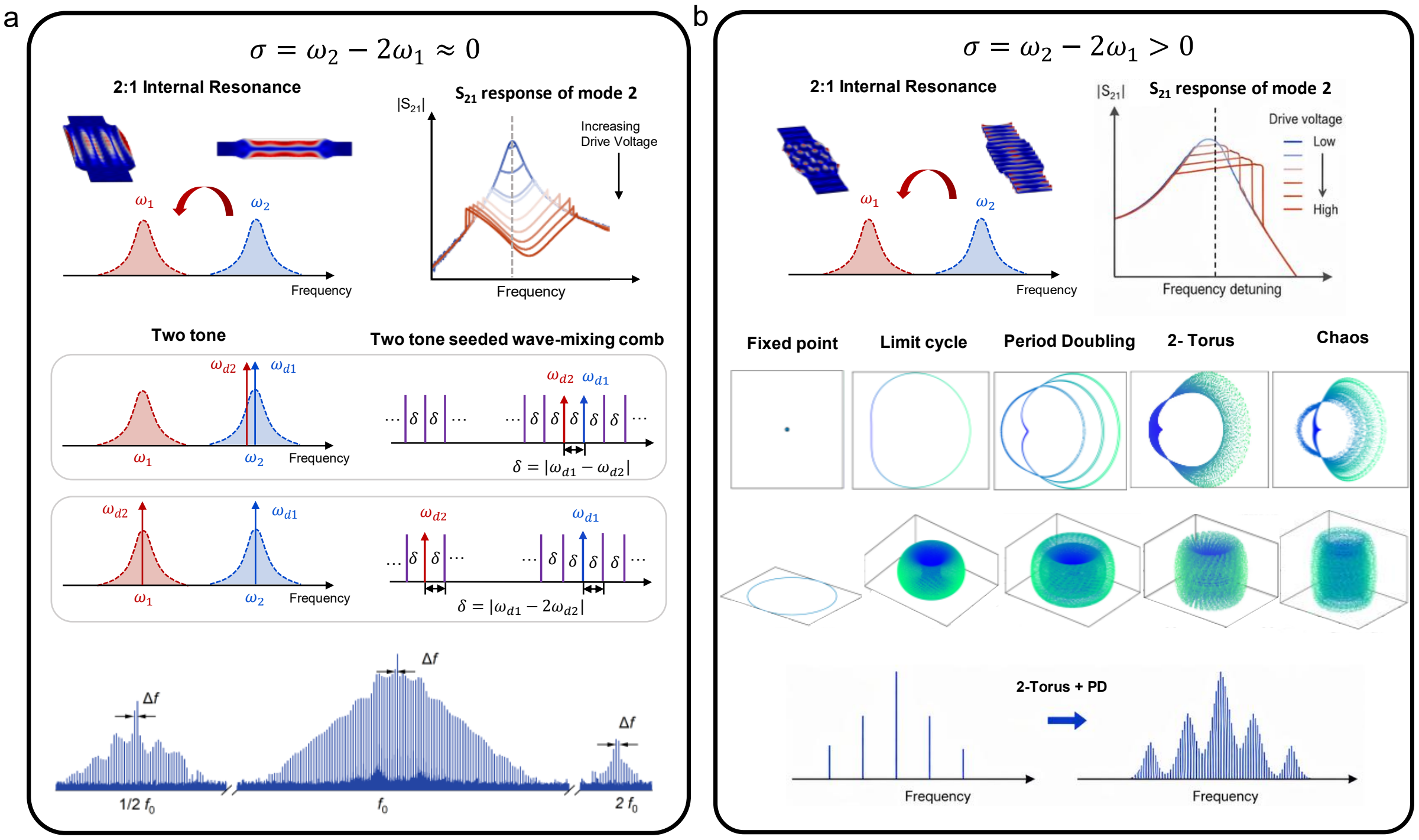

**Fig. 1 | Phononic frequency combs as a multi-clock resource, selected by internal-resonance detuning in HZO CMOS-oxide resonators. a**, Nearly exact 2:1 ratio ($\sigma_i = \omega_2 - 2\omega_1 \approx 0$). Internal resonance between the width-extensional mode $f_2$ and a flexural subharmonic mode $f_1$ produces the characteristic M-shaped resonance in the $S_{21}$ response of $f_2$. Two pump tones ($f_{d1}$, $f_{d2}$) applied near $f_2$ and $f_1$ activate 2:1-assisted four-wave mixing, generating cascaded comb lines around $f_2$ and its subharmonic $f_2/2 \approx f_1$, with spacing set by $\Delta f = |f_{d1} - f_{d2}|$ or $|f_{d1} - 2f_{d2}|$ depending on tone placement. **b**, Detuned 2:1 ratio ($\sigma_i > 0$). Internal resonance between two flexural modes, combined with the frequency mismatch, drives successive dynamical transitions under single-tone excitation — fixed point, limit cycle, 1-torus, 2-torus, period doubling and chaos — as captured in the phase portraits. Period doubling acting on the 2-torus state successively halves the comb spacing, increasing line density without sacrificing spectral span.

The internal frequency mismatch from exact 2:1 commensurability thereby acts as a mechanism selector, and the two pairs yield the two comb families this article compares. Where the Hopf bifurcation is absent ($\sigma_i \approx 0$), comb generation requires two pump tones[53], whose beat seeds cascaded mode-mixing-mediated four-wave mixing, producing more than 170 lines at uniform spacing $\Delta f$, with mutual coherence verified for representative pump and generated lines spanning two octaves around $f_2$, $f_2/2$ and $2f_2$, with a spacing programmed electronically by the pump separation. Where the bifurcation is present ($\sigma_i > 0$), a single pump drives the system through an autonomous sequence of Hopf, torus, and period-doubling bifurcations, producing hierarchical combs of more than 200 lines over a bandwidth reaching 9.5 MHz, with a spacing that is an emergent frequency of the dynamics itself. The coexistence of these two pathways on a single platform — selected deterministically by lithographic geometry — establishes HZO as a geometry-programmable nonlinear medium for broadband, high-density PFCs, and it is what enables the controlled comparison of their frequency stability that follows.

## Two-tone-seeded wave-mixing combs via 2:1 internal resonance

In optical microresonators, comb-forming four-wave mixing is supplied by the cubic Kerr nonlinearity;[12,15] in the phononic domain, combs have been generated through intrinsic three-wave mixing[20] and through bifurcations of nonlinear modes.[21,23] The comb reported here follows a distinct mechanism: two pump tones seed second-order interactions via quadratic coupling, and the generated frequency components repeatedly re-enter subsequent mixing processes to form a cascaded spectral ladder. Quadratic InRes coupling therefore plays, in the acoustic domain, the comb-forming effective four-wave mixing role that Kerr mixing plays in optics.

**Fig. 2a** shows a false-color scanning electron micrograph of the high-aspect-ratio device (width 38 μm, length 200 μm), piezoelectrically actuated through an excitation port and characterized by both electrical and optical readout. Narrow tethers suppress anchor loss and sustain the high quality factors required for strong intermodal coupling at low drive.[39] **Fig. 2b** presents the drive-dependent amplitude response of the fundamental width-extensional mode near $f_2 = 76.8\,\mathrm{MHz}$. Below -20 dBm (22.36 $\mathrm{mV_{rms}}$) the resonance is a symmetric Lorentzian, indicating linear response; above approximately -20 dBm, it evolves into an M-shaped response nearly symmetric about $f_2$. Comparison of the measured frequency response with the ROM (**Supplementary Information 1 S1-2**) confirms that the response over this operating range is governed primarily by the quadratic coupling terms $\alpha u_1 u_2$ and $\beta u_1^2$, with negligible contributions from the cubic nonlinearities (i.e., $\chi_1 \approx 0, \chi_3 \approx 0, \kappa_2 \approx 0$). The clear amplitude saturation at the resonance center is a signature of 2:1 InRes, indicating that the driven high-frequency mode no longer behaves in isolation and intermodal energy transfer has switched on.[45,50,51]

Spectral measurements identify the recipient of that energy (**Supplementary Information 2 S1.1**). Above the same threshold, a component emerges at exactly one-half of the pump frequency. Sweeping the pump across the resonance maps a bounded instability region for this internally excited subharmonic that follows the backbone curve extracted from **Fig. 2b** — an Arnold tongue characteristic of 2:1 InRes, with onset and disappearance in agreement with the theoretical instability boundaries[33] (**Fig. 2c**). Finite-element simulation reveals a higher-order flexural mode whose frequency satisfies the commensurability condition, and optically measured displacement profiles under both external and internal excitation confirm the subharmonic as this flexural mode (**Supplementary Information 2 S1.2**).

The activation of this pair at such low drive follows from how the platform satisfies the three requirements of 2:1 InRes — near-commensurability of the mode frequencies, a quadratic coupling coefficient permitted by the modal symmetries and sufficient spatial overlap of the interacting strain fields[14,44]. In perfectly symmetric doubly clamped beams, the quadratic coupling is weak and appears only through symmetry breaking;[14] here, the large static curvature of the ultrathin beam after release breaks the out-of-plane symmetry and introduces a pronounced quadratic term in the restoring force, which the electrostrictive nonlinearity of the ferroelectric layer further reinforces, collectively lowering the coupling threshold well below that of conventional piezoelectric platforms.[28,38,53] Mechanistically, the interaction is strain-mediated and bidirectional: the in-plane strain of the width-extensional mode dynamically modulates the axial tension seen by the flexural mode through the clamped boundaries and couples to out-of-plane deformation through the Poisson effect, while the activated flexural motion generates mid-plane stretching — the clamped ends preventing free elongation — whose dynamic axial stress perturbs the in-plane stiffness of the width-extensional mode, closing the loop that sustains efficient energy exchange in the internally resonant state.

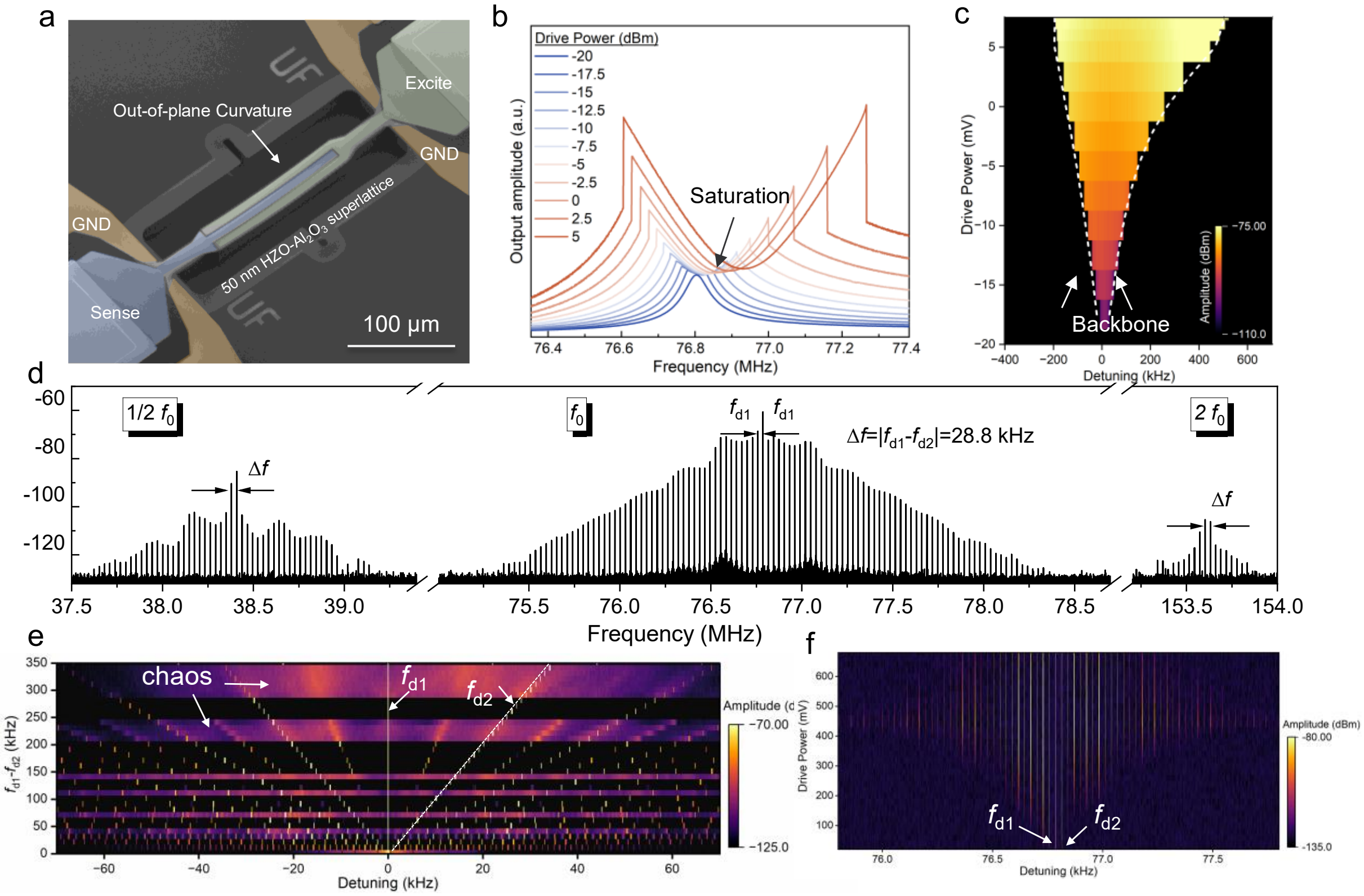

**Fig. 2 | Seeded four-wave-mixing comb generation via 2:1 internal resonance in a high-aspect-ratio HZO CMOS-oxide resonator. a,** False-color SEM image of the doubly clamped beam (width 38 μm, length 200 μm); narrow tethers suppress anchor loss and sustain high-quality-factor modes. **b**, Drive-voltage-dependent amplitude response of the width-extensional mode near $f_2 = 76.8$ MHz, showing the transition from a symmetric Lorentzian to an M-shaped response with amplitude saturation, characteristic of 2:1 InRes energy transfer. **c**, Electrically measured amplitude of the internally driven subharmonic as the pump is swept across the width-extensional resonance $f_2$, forming an Arnold-tongue instability region; the white dashed curve is the backbone extracted from **b**. **d**, Frequency spectrum under two-tone excitation with both pumps near $f_2$ and

$\Delta f = |f_{d1} - f_{d2}| = 28.8$ kHz: approximately 100 comb lines near $f_2$, approximately 50 near $f_2/2$ and approximately 10 near $2f_2$ from electrostrictive harmonic generation — more than 170 lines at uniform spacing $\Delta f$, with mutual coherence verified for representative pump and generated lines, spanning two octaves. **e**, Spectral evolution with $f_{d1}$ fixed and $f_{d2}$ swept, showing the comb spacing locked to $\Delta f$ across a broad detuning range with spacing halved via half-index mixing and chaotic transitions in selected windows. **f**, Power-dependent response with both pump frequencies fixed and the power of $f_{d2}$ varied: the spacing is invariant while the bandwidth is maximized under balanced pumping and reduced under imbalance.

This internally resonant pair constitutes not only the nonlinear medium for comb generation but also a self-reinforcing pathway for cascaded mixing, sustaining higher-order comb lines and extending the comb bandwidth. With two pump tones, $f_{d1}$ and $f_{d2}$, applied near $f_2$, their linear superposition produces a beat at $\Delta f = |f_{d1} - f_{d2}|$. Above the Arnold-tongue threshold, quadratic intermodal coupling converts this beat modulation into new spectral components, which repeatedly re-enter subsequent mixing processes to form a cascaded spectral ladder with free spectral range (FSR) $\Delta f$. This cascade can be seen by expressing the internally resonant responses as

$$u_1 = \sum_n A_n e^{i(\omega_{d1}/2 + n\Delta\omega)t} + c.c., u_2 = \sum_n B_n e^{i(\omega_{d1} + n\Delta\omega)t} + c.c.$$

where $\omega_{di} = 2\pi f_{di}$, $\Delta\omega = 2\pi\Delta f$, and c.c. denotes complex conjugate. The slow flow analysis shows that the quadratic coupling generates convolution among these spectral components of $A_n$ and $B_n$. For example, $B_1 A_0^* \rightarrow A_1$, $B_0 A_1^* \rightarrow A_{-1}$, $A_1 A_1 \rightarrow B_2$. The newly generated $A_n$ and $B_n$ repeatedly re-enter the same interactions, transforming the initial two-tone beat into a self-reinforcing spectral lattice (**Supplementary Information 1 S2.1**). The M-shaped resonance plays a critical role by replacing the narrow, central Lorentzian peak with two high-amplitude off-center branches, enlarging the effective frequency window over which newly generated components away from the resonance center receive strong resonant enhancement, helping the cascade survive to higher orders. For $\Delta f = 28.8$ kHz (**Fig. 2d**), approximately 100 comb lines are generated around $f_0$ and approximately 50 around the subharmonic $f_2/2$, while electrostrictive harmonic generation extends the spectrum to $2f_2$, where approximately 10 further lines appear — in total more than 170 lines at uniform spacing $\Delta f$, with mutual coherence verified for representative pump and generated lines, spanning two octaves from 35 to 153 MHz.

The comb is electronically programmable. With $f_{d1}$ fixed at $f_2$ and $f_{d2}$ swept over a detuning range from −250 to +350 kHz (**Fig. 2e**; complete dataset in **Supplementary Information 2 S2.1**), the spacing remains locked to $\Delta f = |f_{d1} - f_{d2}|$ over a broad range, demonstrating that the FSR is seeded by the externally imposed beat frequency rather than by an intrinsic resonator timescale. Within selected detuning windows, additional components appear at approximately half the primary spacing. This spacing-halved state is attributed to activation of a second, interleaved cascade lattice through cascading of half-index components (**Supplementary Information 1 S2.3**). As discussed later, frequency stability measurements show that half-index components inherit the stability of the applied pumps, distinguishing the pump-referenced half-index cascade from the free-running period-doubled state arising from a dynamical bifurcation. At several distinct detuning windows, the phase-locked cascade loses stability, producing multiple chaotic regions interspersed with stable comb states. These regions likely arise as different sets of competing integer- and half-index mixing pathways become resonant, leading to irregular intermodal energy exchange. The pump placement itself is a further control knob. Applying one tone near $f_2$ and the other near $f_1 \approx f_2/2$ also generates a comb, with spacing set by $\Delta f = |f_{d1} - 2f_{d2}|$, though of substantially smaller extent (**Supplementary Information 2 S2.2**). Here, the quadratic term $\beta u_1^2$ frequency-doubles the lower-frequency pump, producing an effective seed at $2f_{d2}$ near $f_2$ that mixes with $f_{d1}$ to initiate the same cascade. The resulting comb is smaller because the converted seed at $2f_{d2}$ is weaker than a directly applied pump.

Pump balance completes the control set (**Fig. 2f**). With both pump frequencies fixed and the power of $f_{d2}$ varied, the spacing remains locked to $\Delta f$ throughout, whereas the comb bandwidth depends non-monotonically on the relative pump amplitudes: the broadest comb occurs for comparable pumps, and the number of observable lines falls under strong imbalance. Efficient comb formation thus requires both deep beat-note modulation to drive successively higher-order mixing and optimal internally resonant energy transfer over the internal resonance window. Balanced pumping maximizes the modulation component at $\Delta f$ that seeds the cascade, whereas imbalance suppresses the effective modulation depth and detunes the system from the energy-transfer optimum. Consequently, the outer comb lines weaken and the observable bandwidth contracts.

Together, these results establish the seeded comb as an emergent state of the two-tone-driven InRes: the external beat programs the spacing, strong quadratic coupling mediates a self-reinforcing spectral lattice, and the off-center branches of the M-shaped response help sustain the higher-order cascade over an extended bandwidth. More than 170 lines spanning two octaves (35–153 MHz) are generated at drive amplitudes of tens of millivolts, identifying InRes in this platform as a low-threshold route to broadband, dense and electronically programmable comb generation.

## Hierarchical combs via an autonomous Hopf Bifurcation

The detuned modal pair opens a fundamentally different route to comb generation. As established in the preceding analysis, the comb-generation pathway in 2:1 InRes is governed primarily by the internal frequency mismatch, $\sigma_i$. Near exact commensurability ($\sigma_i \approx 0$), preferential high-to-low-frequency energy transfer in 2:1 InRes establishes strong intermodal locking, compared with the low-to-high-frequency pathway in 1:2 InRes, making the pump-locked fixed point less vulnerable to oscillatory destabilization; consequently, no Hopf bifurcation occurs. A finite $\sigma_i$, however, introduces phase drift that competes with coupling-induced synchronization and modifies the amplitude–phase feedback between the two modes. Our slow-flow stability analysis shows that this mismatch allows a complex-conjugate eigenvalue pair ($\lambda_H$) of the Jacobian to cross the imaginary axis, destabilizing the pump-locked fixed point through a Hopf bifurcation and generating a stable slow-flow limit cycle. This periodic oscillation, whose angular frequency is given by the magnitude of the eigenpair's imaginary part at the Hopf onset $\omega_H = |\mathrm{Im}(\lambda_H)|$, modulates the pumped tone to generate an autonomous frequency comb and determine its comb spacing. Thus, finite internal mismatch is the principal ingredient that opens the autonomous Hopf pathway, while damping, coupling strength, drive amplitude, and external detuning $\sigma_e$ determine its threshold and accessible range. Cubic nonlinearities provide secondary quantitative corrections, particularly at high pumping levels, but are not required to initiate the Hopf instability.

This regime is realized in the low-aspect-ratio device (width 42 μm, length 150 μm) at the higher-frequency flexural mode of the detuned pair, near 66.5 MHz (**Supplementary Information 2 S3.1**). Flexural modes, with their large out-of-plane displacement relative to in-plane bulk acoustic motion, are intrinsically susceptible to geometric cubic nonlinearity through mid-plane stretching; here, those cubic nonlinearities coexist with the quadratic intermodal coupling and the electrostrictive nonlinearity of the transduction layer, enriching the dynamics across the entire existence range of the comb. Consistent with the finite modal mismatch, the amplitude–frequency response under elevated drive develops a non-classical, asymmetric M-shaped distortion with a hardening-like skew (**Fig. 3a**). Eventually, drop- and up-jump transitions produce hysteresis, revealing the presence of Duffing nonlinearity, although it remains too weak to dominate the overall resonance response. The frequency mismatch raises the effective threshold for intermodal coupling and skews the instability boundaries[44], and the identification of the coupled modes follows the methodology of the previous section (**Supplementary Information 2 S3.1-3.2**).

With a single pump applied near the higher flexural mode $f_2$, the system crosses the 2:1 InRes threshold at moderate drive and, above approximately 430 mV ($V_{rms}$, ~5.68 dBm), a sparse comb emerges from single-tone excitation (**Fig. 3b**, Comb I), with lines appearing near $f_2$, $f_2/2$ and $2f_2$. Consistent with the slow-flow analysis, this transition occurs through a Hopf bifurcation, at which the stable fixed point loses stability and gives rise to a stable limit cycle in the amplitude–phase dynamics. In the fast-flow dynamics, the corresponding pump-locked periodic orbit undergoes a torus bifurcation, producing an invariant two-torus. At the Hopf onset, the limit-cycle frequency, and therefore the comb-line spacing ($\Delta f$), is given by the imaginary part of the critical eigenvalue pair $f_H = \omega_H/2\pi = |\mathrm{Im}(\lambda_H)|/2\pi$. Beyond onset, the eigenvalues of the unstable fixed point continue to evolve, but the actual comb spacing shifts from this linearized value as the limit cycle evolves nonlinearly. With further variation in drive amplitude or detuning, the slow-flow limit cycle undergoes a secondary Hopf (torus) bifurcation, introducing a second modulation frequency and generating Comb II **(Fig. 3b**), whose autonomous envelope modulation generates sidebands around each primary line, with their spacing set by the newly emerged modulation frequency. Depending on the trajectory through parameter space, a sequence of period-doubling bifurcations leading toward an irregular state also occurs (**Supplementary Information 1 S1.4**). Relative to the two-tone-seeded comb, this autonomous state has a substantially higher threshold and only limited spacing tunability through pump power and detuning, but markedly larger spectral coverage, reaching 9.5 MHz around the carrier. The restricted tunability reflects its bifurcation origin: the comb emerges only after the pump-locked state loses stability, while its spacing is set by an emergent limit-cycle frequency governed collectively by multiple interacting parameters in a multidimensional nonlinear state space.

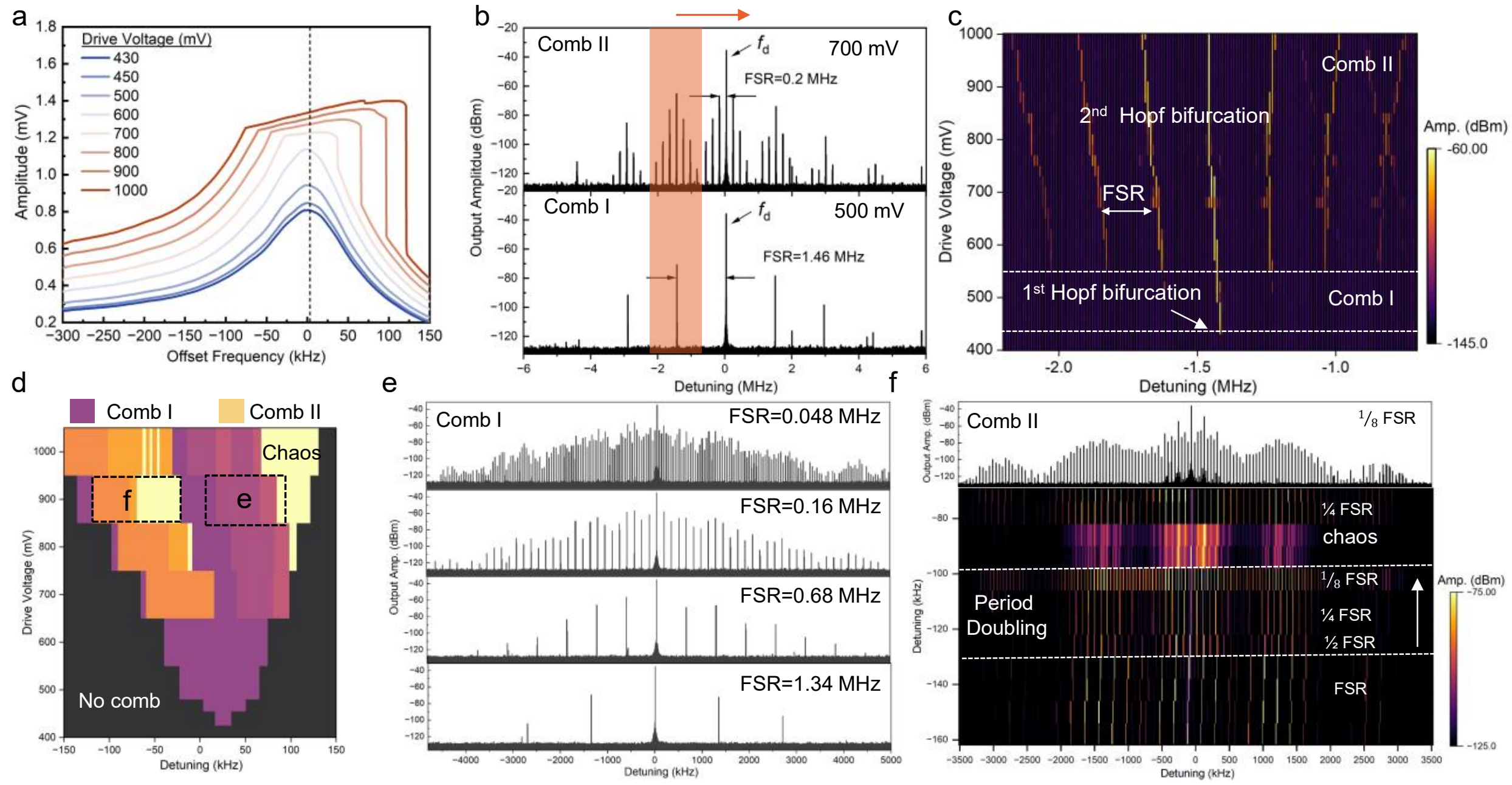

**Fig. 3 | Hierarchical comb generation via coexisting quadratic and cubic nonlinearities in a low-aspect-ratio HZO CMOS-oxide resonator. a**, Drive-voltage-dependent amplitude response of the higher-frequency flexural mode of the detuned pair near 66.5 MHz, showing a non-classical, asymmetric M-shaped distortion with hardening branch only, resulting from the finite frequency mismatch between the interacting flexural modes. **b**, Emergence of the hierarchical comb (Comb II), with two distinct FSRs, from Comb I under single-tone excitation near $f_2$ as the pump power increases from 500 to 700 mV. **c**, Detailed spectral evolution within the highlighted region of **b**: the FSR of Comb I grows with power, and Comb II emerges above approximately 550 mV with a finer spacing exhibiting the same power dependence. **d**, Complete existence map of the comb states versus pump power and frequency: Comb I (purple) and Comb II (yellow) form against the comb-free background (black); darker shading within each region marks period-doubling bifurcations. **e**, Spectral evolution of Comb I along the frequency-sweep column marked in **d**, showing successive free-spectral-range (FSR) halving through period-doubling bifurcations toward chaos. **f**, Colormap of the corresponding evolution of Comb II along its sweep column, showing the same period-doubling cascade and progressive densification without loss of spectral coverage.

Raising the pump power from 500 to 700 mV populates a second family: additional lines emerge around each Comb I line, forming a hierarchical structure with two distinct FSRs (**Fig. 3b**, Comb II). The detailed evolution (**Fig. 3c**) shows the Comb I spacing growing gradually with power after the Hopf bifurcation — shifting the generated tones toward lower frequency — and, above approximately 550 mV, a Comb II appears with substantially finer spacing that exhibits the same power dependence. Time-domain analysis (**Supplementary Information 2 S3.3**) resolves two superimposed envelope modulations in the hierarchical state, with repetition rates matching the inverses of the two spacings, whereas Comb I carries a single envelope — confirming that Comb II arises when the limit cycle born at the Hopf bifurcation itself undergoes a secondary Hopf (torus) instability, producing a two-torus attractor and a commensurate secondary ladder. This cascade of bifurcations densifies the spectrum without sacrificing its coverage.

A complete existence map of the comb states (**Fig. 3d**) was acquired by stepping the pump power and sweeping the pump frequency from low to high at each level. The map confirms that both comb states emerge from the same intermodal-coupling instability, with Comb II confined to higher powers, consistent with the energy required for the secondary bifurcation. Because the map was acquired along a fixed sweep direction, its boundaries carry path-dependent offsets attributable to hysteresis of the underlying nonlinear response. Within both comb regions, fine tuning of the pump frequency drives a progressive halving of the FSR through cascaded period-doubling bifurcations (darker shading in **Fig. 3d**).

This cascade is examined in detail in **Figs. 3e** and **3f**, which follow the frequency-sweep column marked in **Fig. 3d**: **Fig. 3e** presents successive Comb I spectra at discrete pump frequencies, and **Fig. 3f** displays the corresponding evolution of Comb II as a continuous colormap. Both families exhibit a clean sequence of free-spectral-range halvings with increasing pump frequency — a period-doubling route toward chaos — that interleaves new lines while preserving the outer spectral envelope, enabling densification without loss of coverage, in contrast to the seeded comb, whose span contracts when its spacing is reduced (**Supplementary Information 2 S2.3**). At the deepest measured period-doubled state — a drive of 1.2 V detuned 90–100 kHz from $f_0$ (**Fig. 3e**, top panel) — the $f_0$ cluster alone contains approximately 170 lines across its 9.5-MHz span, and through the combined action of the initial Hopf bifurcation, the secondary Hopf bifurcation and the period-doubling cascade, more than 200 comb lines are generated near $f_0$. The comb extends to subharmonic and harmonic clusters near $f_0/2$ and $2f_0$, whose spectra at drive conditions matched to the $f_0$-cluster data are presented in **Supplementary Information 2 S3.2**.

## Scaling comb generation to gigahertz frequencies

Because the InRes condition requires only frequency commensurability and sufficient spatial overlap between the strain fields of the interacting modes, it scales. The strategy is to preserve the known coupled mode-shape pair while shrinking the geometry, so that the commensurability condition is carried to the target frequency without altering the modal symmetries or the overlap integral. The dimensional control of the ALD-grown stack provides a direct route: the internal-resonance frequency is tuned independently through the stack thickness and the in-plane dimensions, finite-element simulations confirm that the condition is preserved across a broad range of scaled designs, and a ±5% geometric compensation applied to each design absorbs fabrication-induced offsets across the fabricated device population.

Among the fabricated designs, an in-plane scale factor of 0.2 applied to the 38 μm × 200 μm reference geometry, together with a 15-nm HZO–$Al_2O_3$ stack, produced the highest-frequency device. Its third width-extensional mode at 1.07 GHz participates in a non-degenerate combination resonance with lower-frequency flexural modes at $f_1 = 441$ MHz and $f_2 = 625$ MHz, satisfying the condition $f_3 \approx f_1 + f_2$. This intermodal interaction via combination resonance represents the non-degenerate counterpart of 2:1 InRes. The three-mode interaction is described by the cubic potential $V = \kappa u_1 u_2 u_3$, whose derivatives generate the quadratic coupling terms ($u_2 u_3$, $u_1 u_3$, and $u_1 u_2$) in the respective modal equations. Here, $u_1 u_2$ drives the sum-frequency ($f_1 + f_2$), while $u_2 u_3$ and $u_1 u_3$ provide reciprocal difference-frequency pathways ($f_3 - f_2 \approx f_1$, $f_3 - f_1 \approx f_2$), together enabling the bidirectional three-wave cascade mixing $f_1 + f_2 \Leftrightarrow f_3$. Thus, two distinct lower-frequency modes replace the frequency-degenerate subharmonic pair of 2:1 InRes, while preserving the quadratic mixing cascade to gigahertz comb generation. **Figure 4a-b** compares the reference and scaled device geometries in scanning electron micrographs. The thickness of the $SiO_2$ layers were co-scaled to preserve the temperature-compensation condition of the width-extensional mode. Under two-tone excitation near the third width-extensional mode at approximately 1 GHz, quadratic intermodal interactions drive a cascaded mixing process across the combination-resonant mode network. The pump separation sets the FSR, while successive mixing among the participating modes generates line clusters near $f_1$, $f_2$, $f_3$, and $f_3 + f_2$, while electrostrictive harmonic generation further extends this cascade to $2f_1$ and $2f_3$, producing comb clusters distributed from approximately 0.44 to 2.1 GHz (**Fig. 4c**).

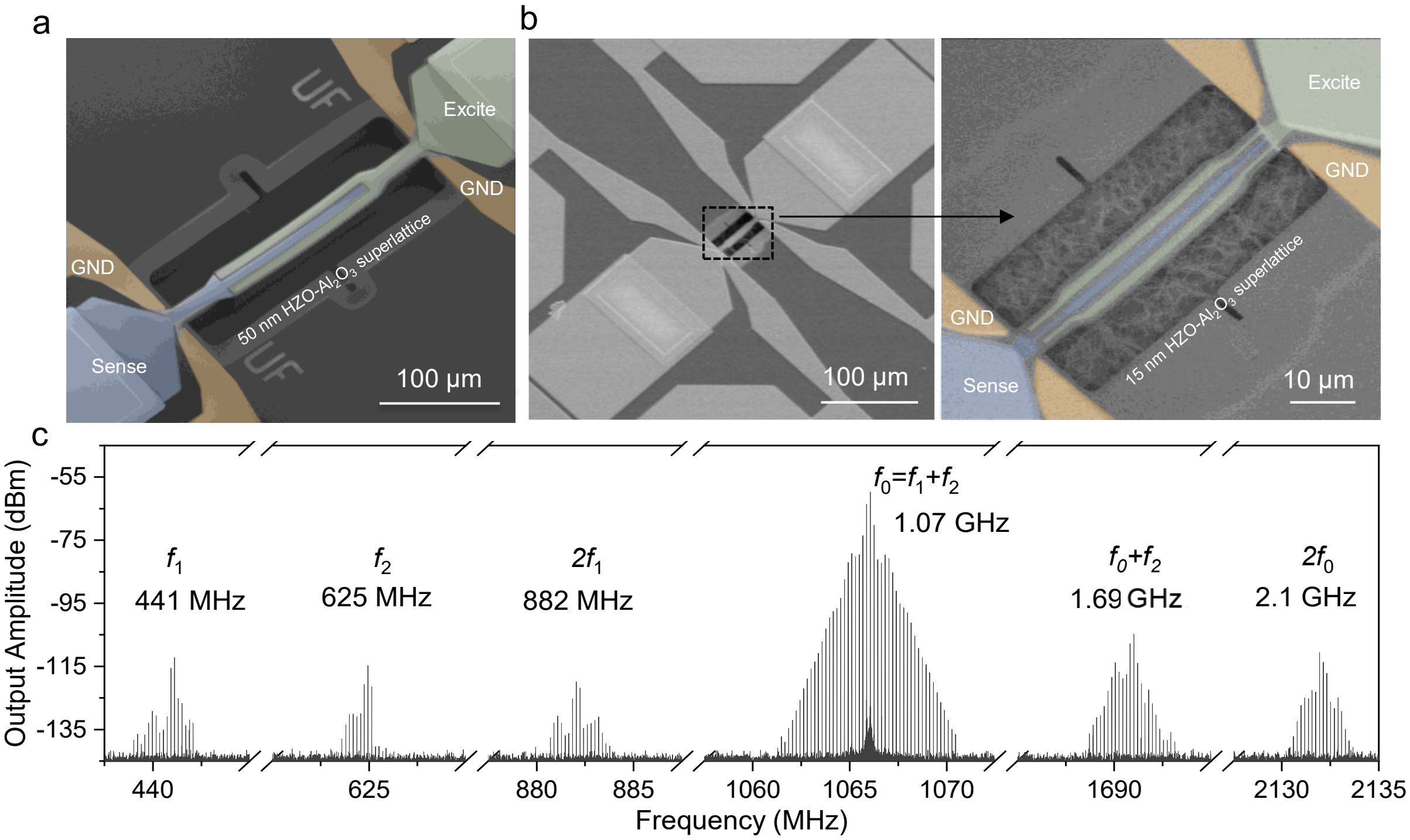

**Fig. 4 | Gigahertz-scaled comb generation enabled by the dimensional control of the HZO CMOS-oxide stack. a,b**, False-color SEM images comparing the reference geometry **a** with the 0.2-scaled device and a magnified view of the scaled beam **b**. **c**, Measured comb spectrum under two-tone excitation near the third width-extensional mode at $f_0 \approx 1.07$ GHz, with clusters at $f_1 = 441$ MHz, $f_2 = 625$ MHz, $2f_1 = 882$ MHz, $f_0 = f_1 + f_2$, $f_0 + f_2 = 1.69$ GHz and $2f_0 = 2.1$ GHz— comb clusters distributed from approximately 0.44 to 2.1 GHz with FSR set by the pump-tone separation $\Delta f = |f_{d1} - f_{d2}|$.

The point of this scalability is not to reproduce the terahertz bandwidths of optical combs; it is to translate the comb principle into the frequency range and the material stack in which electronic systems actually operate. Gigahertz operation places comb lines directly in the bands of wireless transceivers, data converters and radar front ends, and electrical coupling between individually scaled resonators offers a route to composite combs whose bandwidth exceeds that of any single device — removing the coverage constraint that has historically confined phononic combs to demonstrations.

## Mechanism-limited short-term stability

For optical combs, the property that converted spectra into infrastructure was not line count but the coherence and stability of the lines, and their evolution across operating regimes became the standard diagnostic of the underlying dynamics.[54] Comparable investigations of phononic combs remain scarce[35]. Here we track the MDEV of individual lines, by heterodyne down-conversion against a stable reference, across the full ladder of nonlinear states supported by one platform: the linearly driven resonance, the internally resonant subharmonic, the seeded wave-mixing comb and the autonomous Hopf comb.

The organizing principle is a hierarchy of phase constraints. Two tones are mutually coherent when their relative phase evolves deterministically so that their beat note stays spectrally sharp.[55] They are phase-locked when an integer combination $q\varphi_a - p\varphi_b$ of their phases settles to a constant. Phase locking therefore implies mutual coherence, whereas mutual coherence alone does not necessarily imply phase locking. The generation routes studied here differ in the number and origin of the independent phase degrees of freedom that remain unconstrained. In a 2:1 InRes system, the subharmonic sits at $\omega_d/2$ and is phase locked to the pump through the condition $2\phi_s - \phi_d = \text{const}$, up to the intrinsic $\pi$ ambiguity. In a two-tone seeded wave-mixing comb, each line carries $\varphi_1 + n\varphi_r$ with $\varphi_r = \varphi_1 - \varphi_2$ fixed by the pumps, leaving no additional phase degree of freedom. In contrast, the autonomous comb carries exactly one free parameter: its spacing $\Omega$ is an emergent limit-cycle frequency, so the line phases take the form $\varphi_d + n\varphi_H$, in which the carrier component is pinned to the drive while $\varphi_H$ free-runs — the comb remains a rigid, mutually coherent two-parameter object whose offset is referenced to nothing external. Noise drives diffusion most strongly along unconstrained phase degrees of freedom, and this ledger identifies where such diffusion can occur.

For each state, the composite output —a fast carrier with a slow amplitude envelope whenever a comb is present (**Fig. 5a,b**)—is mixed with an independently referenced comparison tone to produce a beat near 5 kHz (**Fig. 5c**), and the resulting timestamp record is used to calculate the carrier-referred MDEV of the selected line (**Methods, Supplementary Information 3**). Before introducing the resonator, the reference and readout chain was calibrated using common-reference and independent-reference heterodyne measurements. The common-reference measurement characterizes the non-common residual instability of the synthesis and readout paths, whereas the independent two-Rakon comparison contains the relative fluctuations of the two independently referenced synthesis paths together with the residual measurement-chain contribution. The latter is reported directly, without de-embedding either oscillator, and serves as the practical reference and readout background in **Fig. 5d**. Both calibration traces exhibit an approximately $\tau^{-3/2}$ dependence over much of the short-averaging-time interval, consistent with white-phase-modulation-like behavior, and the independent-reference trace approaches a flicker-frequency plateau near $10^3$ s. Because the same short-term dependence is observed in the absence of the resonator, its appearance in a resonator-generated tone indicates preservation of the measured stability class but cannot, by itself, be assigned uniquely to the reference oscillator, the resonator or the readout chain.

Against this floor, every externally referenced state behaves as a passive element of the synthesis chain. The linearly driven resonance, the 2:1 subharmonic and the seeded wave-mixing comb all preserve the $\tau^{-3/2}$ scaling of the source, differing only in level (**Fig. 5d**). The small offset of the linear resonance is attributed to thermomechanical motion and additive measurement noise, whereas the approximately one-order higher instability of the subharmonic arises from its lower signal-to-noise ratio (SNR), as the electrode geometry is not optimized for transduction at $f_2/2$. Consistent with this interpretation, raising the drive power increases the subharmonic amplitude while monotonically reducing its MDEV (**Supplementary Information 2 Fig. S4.4**), revealing a nearly exponential relationship between line amplitude and measured instability that identifies the degradation as SNR-limited rather than intrinsic. Although the nonlinearity synthesizes new frequencies, the resonator does not become an autonomous oscillator in these externally referenced states: each state inherits the phase of the pump through forced excitation, phase locking or cascaded mixing, and the seeded comb operates, in effect, as a passive mechanical frequency synthesizer.

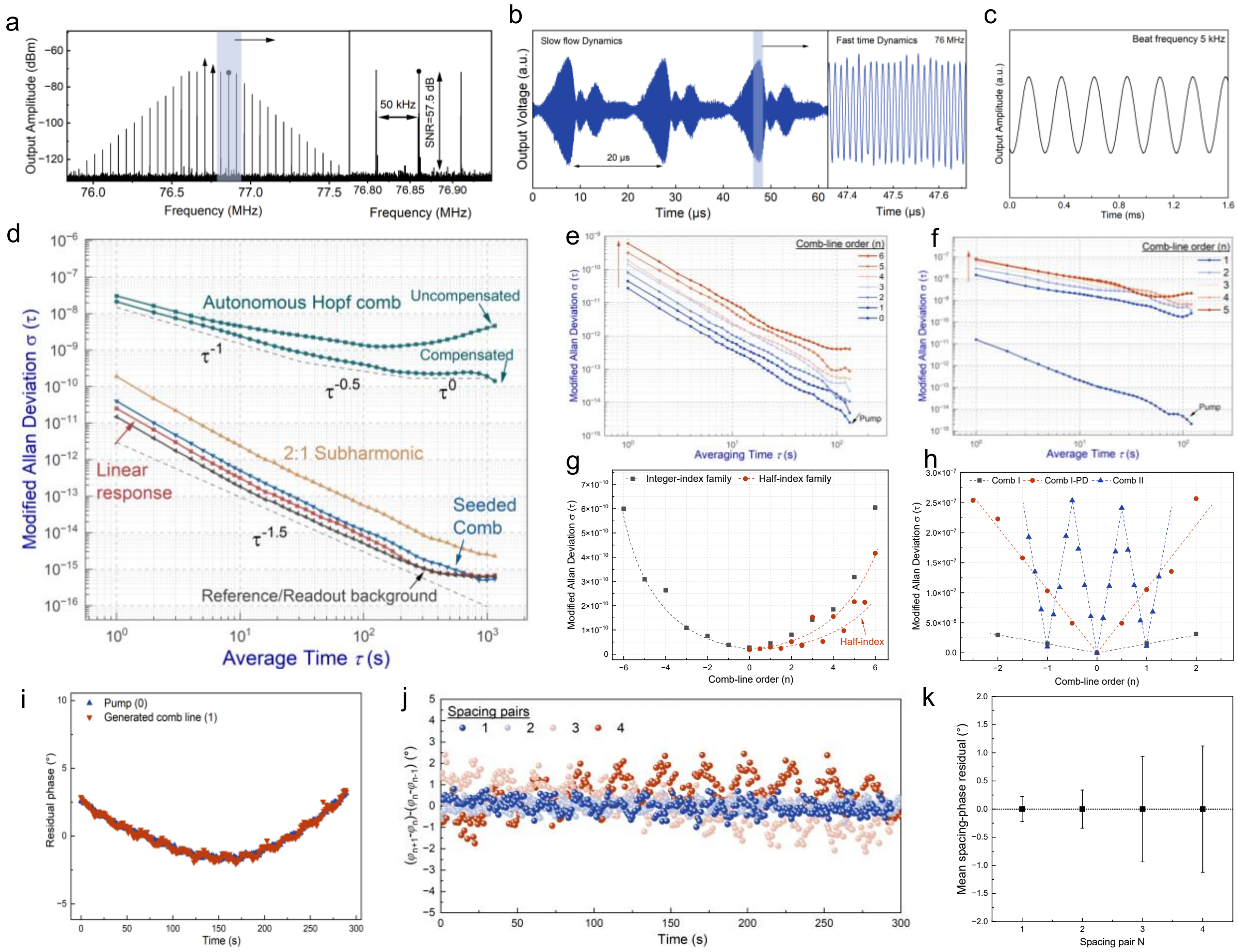

**Fig. 5 | Generation-mechanism dependence of comb-line frequency stability and direct mutual-coherence measurements. a**, Measured spectrum of the two-tone-seeded wave-mixing comb, with the lines selected for stability and phase measurements indicated by arrows; the inset resolves a selected comb line with a signal-to-noise ratio of 57.5 dB. **b**, Time-domain output of the seeded comb showing the slow envelope modulation and, in the magnified interval, the fast carrier oscillation near 76 MHz. **c**, Heterodyne beat near 5 kHz used for timestamp-based frequency

counting. **d**, Carrier-referred modified Allan deviation (MDEV) of the reference and readout background, linearly driven response, internally resonant 2:1 subharmonic, seeded wave-mixing comb, autonomous Hopf comb, and autonomous combs measured in compensated and uncompensated resonators. Dashed guides indicate the canonical MDEV dependences associated with white phase modulation ($\tau^{-3/2}$), flicker phase modulation ($\tau^{-1}$), white frequency modulation ($\tau^{-1/2}$) and flicker frequency modulation ($\tau^{0}$). The externally referenced states remain close to the white-phase-modulation-like reference and readout background, whereas the autonomous states exhibit higher instability and different averaging-time dependences. **e**,**f**, MDEV as a function of averaging time for selected comb-line orders of the seeded wave-mixing comb (e) and autonomous Hopf comb (f); the corresponding pump trace is shown for comparison. **g**, MDEV at $\tau = 1$ s as a function of signed comb-line order for the integer-index and interleaved half-index families of the two-tone-seeded comb. The dashed curves show the amplitude-dependent trends associated with the measured line strengths. **h**, MDEV at $\tau = 1$ s as a function of comb-line order for Comb I, period-doubled Comb I and hierarchical Comb II. For Comb II, the line order is indexed using the larger Comb I spacing. **i**, Residual phases of the pump ($n = 0$) and a generated comb line ($n = 1$) over a 300-s record after removal of their deterministic linear phase evolution. **j**, Differential spacing-phase residual, $(\phi_{n+1} - \phi_n) - (\phi_n - \phi_{n-1})$, for four adjacent spacing pairs. **k**, Mean differential spacing-phase residual for each measured spacing pair; error bars denote one standard deviation over the 300-s record. The bounded residuals and near-zero means support mutual coherence and spacing rigidity among the measured pump and generated lines.

This inheritance has a quantitative form that mirrors the multiplication problem of the opening paragraph. Just as a PLL multiplying by $N$ raises reference phase noise by $20 \log N$, a comb transfers fluctuations of its spacing with a weight that increases with line order. For $f_{c,n} = f_d + n\Delta f$, the spacing contribution scales approximately with $|n|$ when spacing fluctuations dominate and covariance with the carrier is negligible.[7,35] The comb does not evade the multiplication law — it determines what is multiplied. Within each family the MDEV traces remain nearly parallel while rising with order (**Fig. 5e,f**), but the two families rise for opposite reasons. In the seeded comb the spacing is the difference of two tones generated by separate synthesizers referenced to the same clock, so spacing noise $\delta\varphi_r$ is suppressed by common-mode rejection. The measured MDEV increases approximately exponentially with order, tracking the decaying amplitude of higher-order lines, and decreases monotonically with increasing drive power **(Supplementary Information 2 Fig. S4.5**), indicating that SNR dominates the measured stability, while the intrinsic $n\sigma_1$ contribution remains present (**Fig. 5e,g**). In the Hopf comb the spacing is selected by the autonomous nonlinear dynamics, and the associated phase fluctuations $\delta\varphi_H$ are sufficiently large to dominate the measured instability. The MDEV therefore grows approximately linearly with order (**Fig. 5h**) and depends non-monotonically on drive power (**Supplementary Information 2 Fig. S4.6**), the signatures of intrinsic nonlinear noise. Although SNR is not intrinsic to the nonlinear dynamics, it sets the number of usable lines above the detection floor and is therefore a practical design quantity in its own right.

The Hopf comb pays for its emergent spacing at every averaging time. Upon crossing the Hopf threshold, the stable phase-locked fixed point loses stability through a Hopf bifurcation, giving rise to an autonomous limit cycle with a phase that is no longer constrained by the external reference. The MDEV consequently departs from the source in both level and law (**Fig. 5d**): the instability rises from approximately $10^{-11}$ to $10^{-8}$ at one-second averaging time, nearly three orders of magnitude, and the $\tau^{-3/2}$ scaling gives way to a slope near $\tau^{-1}$ (flicker phase noise) at short times, $\tau^{-1/2}$ (white frequency noise) beyond roughly 20 s and a plateau (flicker frequency noise) near 80 s, marking a progressive transition from pump-referenced dynamics to the canonical noise progression of a free-running oscillator[56,57]. The normal form of the Hopf bifurcation[37],

$$\dot{z} = (\mu + i\omega)z - (1 + i\beta)|z|^2 z$$

provides the theoretical framework for interpreting this transition. In the locked states, amplitude perturbations barely touch the phase, which is continuously corrected by the external reference. Once the system undergoes a Hopf bifurcation, the non-isochronicity $\beta$ couples amplitude fluctuations, including thermomechanical and environmental perturbations, into phase fluctuations, leading to phase diffusion of the autonomous phase $\varphi_H$. Stochastic analyses of this normal form establish the conditions under which a noisy limit cycle retains coherence[37]. The measurements presented here supply what such analyses have lacked — the transfer from generation mechanism to a measurable stability metric — the role played for soliton microcombs by quantum-diffusion theory.[36]

Stability also reads out the bifurcation sequence directly. For the hierarchical comb, the MDEV separates into two branches rather than extending Comb I continuously (**Fig. 5h**): lines inherited from Comb I retain low instability, while the interleaved family is substantially noisier — a stability signature consistent with the second family being born at a secondary Hopf bifurcation and acquiring additional phase noise there, analogous to primary Kerr comb lines seed noisier subcombs in optical microresonators.[54] The separated noise branches thereby support its identification as hierarchical, rather than uniformly spacing-reduced, by its noise.

The most stringent demonstration comes when spectrally similar phenomena arise from fundamentally different mechanisms. In the spacing-halved two-tone seeded wave-mixing comb, the newly generated half-index lines ($f_{d1} + n\Delta f/2$ for odd $n$) exhibit higher frequency stability than their parent integer-spacing family ($f_{d1} + n\Delta f$), despite their lower SNR (**Fig. 5g**). This result reflects coherent phase inheritance through the nonlinear wave-mixing process rather than a measurement artifact. A plausible pathway, consistent with the phase ledger, involves simultaneous subharmonic locking of both pumps: producing internally generated components $f_{d1}/2$ and $f_{d2}/2$, separated by $\Delta f/2$. Their mutual mixing and subsequent cascading populate the interleaved half-spacing lattice without introducing a free phase (**Supplementary Information 2 S4.7**). In contrast, the spacing-halved Hopf comb, reached by period doubling of the autonomous torus on the route to chaos, instead exhibits systematically degraded short-term frequency stability and substantially higher MDEV (**Fig. 5h**). The same spacing-halving phenomenon can arise from fundamentally different nonlinear mechanisms. Their physical origins are distinguished not by the spectrum, but by the short-term frequency stability.

Mutual coherence of the two-tone seeded wave-mixing comb is measured directly by simultaneous phase tracking. Pairs of lines are down-converted and their phases digitized simultaneously on two lock-in demodulators sharing one reference (**Methods, Supplementary Information 2 Fig. S5.1**). After removal of the deterministic linear phase evolution associated with their frequency difference, the residual phases exhibit the same temporal drift (**Fig. 5i**), and pairwise subtraction leaves a zero-mean residual bounded within $\pm 0.4°$ with a stationary envelope (**Supplementary Information 2 Fig. S5.2**) — the bounded, non-diffusive relative phase supports mutual coherence between the pump and the measured generated lines. The residual standard deviation increases from 0.15° at $n = 1$ to 0.66° at $n = 4$, consistent with the reduced signal-to-noise ratio of the higher-order lines. Spacing coherence is evaluated from the second phase difference $(\varphi_{n+1} - \varphi_n) - (\varphi_n - \varphi_{n-1})$, which eliminates the deterministic linear phase evolution without fitting. Any line-dependent repetition rate would produce a residual slope, whereas none is observed (**Fig. 5j-k**).

## Material-limited long-term stability

At averaging times beyond tens of seconds a different variable takes over: temperature. Its effect, however, depends on the comb-generation mechanism. In a two-tone seeded wave-mixing comb, externally fixed pump frequencies largely determine the comb grid, while temperature-dependent shifts of the participating mechanical modes mainly modify pump–mode detuning, internal-resonance matching, conversion efficiency, and therefore the observable comb bandwidth and amplitude distribution (**Supplementary Information 2 S6.1**). In contrast, the single-pump Hopf comb exhibits an additional sensitivity arising from the nonlinear limit-cycle dynamics. Temperature-induced perturbations of the operating point can be converted into phase fluctuations through amplitude-to-phase (AM-to-PM) coupling, giving rise to distinct long-term stability characteristics.

The single-pump comb used for the long-term measurements is generated on an internal-resonance pair between two out-of-plane flexural modes near 68 MHz (the width-extensional mode of the same device also supports a Hopf comb, of smaller spectral extent); because the $SiO_2$ layer was designed primarily to compensate the width-extensional family[41], we measured the TCF of the approximately 68 MHz flexural mode directly (**Methods**), obtaining $TCF_1 = -8.1\ ppm/°C$ (**Supplementary Information 2, Fig. S6.2**). This magnitude is approximately ninefold smaller than the $TCF_1 \approx -74\ ppm/°C$ reported for uncompensated HZO resonators[41], indicating substantial partial compensation even for a flexural mode not targeted by the original design. Measured in air, without vacuum packaging or active thermal control, the comb lines of the compensated HZO–$Al_2O_3$–$SiO_2$ device show no random-walk upturn across the full measured range, out to $10^3$ s, settling instead on a flicker-frequency floor near $2\times10^{-10}$ (**Fig. 5d**) — long-term behavior otherwise obtained commonly inside ovenized or actively corrected oscillators. An otherwise identical uncompensated HZO resonator, driven into the same flexural-pair comb state, develops the $\tau^{+1/2}$ signature of random-walk frequency noise within a few hundred seconds. The same signature appears across materials: an uncompensated aluminum scandium nitride resonator supporting a single-pump comb at 64.255 MHz [46,47] develops random walk beyond approximately $10^2$ s, even though its higher quality factor gives it the lower short-term MDEV — confirming that long-term degradation is set by temperature sensitivity rather than by material choice, and that the short-term mechanism law of the previous section is platform-independent.

The dichotomy between comb families sharpens the design rule. Seeded lines lie at $f_{d1} + n\Delta f$: their frequencies are fixed by the externally supplied pump tones, so temperature compensation primarily stabilizes the operating point and the comb's existence range rather than the line frequencies themselves. Autonomous lines lie at $f_d + nf_H$, with $f_H$ an eigenfrequency of the resonator's nonlinear dynamics. Temperature therefore shifts these lines both through $f_H$ and by perturbing the nonlinear operating point. The compensation layers reduce this sensitivity but do not make the autonomous comb temperature-independent. As a result, the compensated stack suppresses the random-walk upturn observed in uncompensated resonators. Mechanism at short times, material at long times: together, they determine when a phononic comb is an instrument rather than a spectrum.

## Conclusions

We have shown that a single ferroelectric HZO resonator, driven by one or two tones, generates broadband phononic frequency combs: more than 170 lines distributed over two octaves, with mutual coherence verified for representative pump and generated lines in the seeded regime; more than 200 lines in the hierarchical autonomous regime; and comb operation extended to 2.1 GHz. More consequentially, we show that whether such a comb can serve as frequency infrastructure is governed by rules that can be engineered. Three design levers emerge. Lithographic geometry programs the internal-resonance detuning, $\sigma_i = \omega_2 - 2\omega_1$, selecting seeded wave mixing near zero mismatch and the autonomous Hopf cascade at finite mismatch. The generation mechanism sets the short-term stability class through its ledger of free phases: the seeded comb, with no additional free spacing phase, retains the pump-referenced white-phase-noise-like stability class, with a one-second modified Allan deviation (MDEV) near $10^{-11}$, whereas the autonomous comb acquires one free spacing phase and its one-second MDEV reaches approximately $10^{-8}$. The material stack sets the long-term class through the temperature coefficients of the participating modes: in air and without active thermal control, the compensated HZO–$Al_2O_3$–$SiO_2$ stack holds the measured comb lines near a flicker-frequency floor, whereas uncompensated resonators develop random-walk frequency noise. Spectrum-first evaluation of phononic combs is thereby replaced by specification: a comb state can be chosen for a function as any engineered component is chosen—by its class, not by its appearance.

For integrated electronics, the immediate consequence is architectural. The systems that motivated this work—interleaved data converters, phased arrays, multicarrier transceivers, sensor arrays and chiplet fabrics—do not need arbitrary frequencies; they need arithmetic grids of mutually coherent references, and that is precisely what a single comb resonator can supply. Local phase-locked loops locked to nearby comb lines could operate at substantially reduced multiplication ratios, reducing the $20 \log N$ phase-noise multiplication penalty and the synchronization circuitry otherwise required to recover inter-domain coherence loop by loop. The remaining engineering challenges are visible and tractable within the platform: the measured comb-line powers of $-80$ to $-120$ dBm can be addressed through output buffering, higher electromechanical coupling in thicker or higher-polarization stacks, and electrode geometries matched to the subharmonic family[39,40]; and the existence maps convert drive selection from search into design.

Beyond clocking, the comb is a general radiofrequency resource whose reach follows directly from the stability rules established here. Two seeded combs with electronically offset spacings—a configuration directly enabled by control of the pump separation—could form a dual-comb correlator that folds a wide radiofrequency band onto a dense grid of low-frequency beats, enabling parallel channelization, spectral sensing and correlation with reduced reliance on filter banks and high-speed data conversion, in direct analogy to dual-comb spectroscopy and ranging in optics.[16,17] The same mutually coherent carriers could support frequency-multiplexed analog computing[18,58], with two advantages native to this platform: the ferroelectric transducer provides electrical programmability of the spacing, amplitude and device state[39,40], and the medium supplies the required nonlinearity without added components. The autonomous and near-chaotic states may, in turn, furnish deterministic high-dimensional spectral bases for analog correlation and classification, although that regime will demand stability metrics of its own. Every one of these functions inherits the requirement quantified in this work: only comb states certified by mechanism and by material qualify as infrastructure—a comb that cannot be trusted is a spectrum, not an instrument.

Optical frequency combs became technology when their field advanced from generating spectra to certifying their coherence. The platform, mechanisms and stability laws reported here advance that transition in the native frequency domain of electronics—within the CMOS oxide toolbox and with no electro-optic conversion in the loop[19]. A mechanical resonator has long served electronics as its clock; operated as described here, it becomes the clockwork.

## Methods

### HZO-$Al_2O_3$ superlattice deposition

The amorphous HZO–$Al_2O_3$ superlattice was fabricated using a Cambridge NanoTech Veeco Fiji atomic layer deposition (ALD) system operated at a process temperature of 200 °C. Tetrakis(dimethylamido)hafnium(IV) and tetrakis(dimethylamido)zirconium(IV) were used as the Hf and Zr precursors, respectively. After each ALD cycle, a 300 W hydrogen/oxygen plasma treatment was applied to oxidize the precursors and promote the formation of the orthorhombic phase. The nominally 50-nm transduction layer consisted of four repetitions of 9.1 nm HZO/1 nm $Al_2O_3$, followed by a final 9.1-nm HZO layer. The 15-nm transduction layer consisted of 7 nm HZO/1 nm $Al_2O_3$/7 nm HZO.

**$SiO_2$ compensation layer deposition**

Plasma-enhanced chemical vapour deposition (PECVD; Plasmatherm 790) was used to deposit the top and bottom $SiO_2$ passivation and temperature-compensation layers. A nominal $SiO_2$ thickness of 350 nm was used with the 50-nm transduction stack and 105 nm with the 15-nm stack, maintaining a constant $SiO_2$-to-transducer thickness ratio. The $SiO_2$ thickness was verified using a KLA Filmetrics F54 reflectometer.

**NEMS resonator fabrication**

For the $SiO_2$-compensated resonator, a $SiO_2$ layer was first deposited on a high-resistivity Si substrate by PECVD, serving as both the passivation and temperature-compensation layer. A 25 nm tungsten (W) film was then sputtered (Lab 18) and patterned by sulfur hexafluoride ($SF_6$)-based reactive ion etching (RIE) to define the bottom electrodes, followed by the lift-off deposition of 150 nm platinum (Pt) for the bottom routing. The HZO–$Al_2O_3$ superlattice transducer was subsequently deposited by ALD. Afterward, a 20 nm W top electrode was sputtered, and rapid thermal annealing (RTA) was carried out in a JetFirst RTP 150 system at 550 °C for 20 s in forming gas. The top W layer was patterned using RIE, after which a second 150 nm Pt lift-off process was performed to form the top routing. Openings to the top and bottom Pt contact pads were created by ion milling (Nanoquest II Ion Mill), and 200 nm Pt pad plugs were deposited by lift-off to provide low-resistance electrical contacts. Finally, the resonator trenches were defined by ion milling, and the structures were released by top-side silicon etching using xenon difluoride ($XeF_2$). For the uncompensated HZO resonator, a 30 nm HZO passivation layer was first deposited on the high-resistivity Si substrate by ALD to protect the bottom electrodes during the release process. After formation of the resonator stack, a symmetric 30 nm HZO capping layer was deposited by ALD to protect the resonant body during release. Additional fabrication details can be found in our previous work.[38–41].

**SEM and optical microscopy characterization**

The SEM images were acquired using a Hitachi SU8000 cold-field-emission scanning electron microscope.

**Nonlinear characterization**

The transmission response ($S_{21}$) of the resonators was measured using a vector network analyzer (VNA), from which the frequency response curves (FRCs) were obtained under different drive amplitudes. Nonlinear dynamic behaviors, including 2:1 internal resonance (InRes) and frequency comb generation, were characterized by electrically pumping the devices with a signal generator while monitoring the output spectrum using a spectrum analyzer (SA). The pump frequency was tuned in the vicinity of the resonant mode, and the pump power was varied to identify the onset of nonlinear responses and the frequency comb. The spatial vibration profiles of the coupled modes participating in the 2:1 internal resonance were measured using a laser Doppler vibrometer (Polytec MSA-600) to verify their corresponding mode shapes. Time-domain signals were recorded with a digital oscilloscope.

**Frequency-stability measurements.**

Individual resonator-generated tones were heterodyned against independently referenced comparison tones to produce beat notes near 5 kHz. Beat periods were recorded using a Keysight 53230A counter/timer operated in Array Timestamp mode, and carrier-referred MDEV was calculated from the resulting residual-time records using an overlapping estimator without detrending. Rakon clock 1 referenced the resonator drive synthesizer or synthesizers, whereas Rakon clock 2 referenced the comparison synthesizer. Common-reference and independent-reference measurements characterized the practical synthesis and readout background; the independent two-Rakon result was reported directly without de-embedding either oscillator. All stability measurements were performed in air without vacuum packaging or temperature control; additional acquisition and processing details are provided in **Supplementary Information 3**. Mutual-coherence measurements. Selected pairs of seeded-comb lines were simultaneously down-converted, and their phases were digitized using the two input channels of a Zurich Instruments UHFLI sharing a common time base. The deterministic linear phase evolution associated with the frequency offset was removed before calculating the differential residual phase over the 300-s records.

**Temperature-coefficient characterization.**

The temperature coefficient of frequency (TCF) of the devices was characterized by locking the resonator mode to a Zurich Instruments UHFLI while performing temperature sweeps. A Nextron temperature-controlled chamber was used to conduct both cooling and heating cycles at a rate of 2.5 °C/min. Prior to the measurements, the samples were held at 100 °C for 1 hour to relieve residual stress and stabilize the resonance frequency. The temperature was then swept downward and upward to record the frequency response. No measurable hysteresis was observed between the cooling and heating cycles. The first-order temperature coefficient of frequency ($TCF_1$) was extracted by fitting the frequency–temperature data with a second-order polynomial.

**Slow-flow analysis.**

The coupled modal equations were solved using the method of multiple scales under weak damping, weak nonlinearity, and near-2:1 internal resonance. Steady-state responses were obtained by numerically solving the resulting nonlinear slow-flow equations over the prescribed excitation-detuning and forcing ranges. Their local stability and the saddle-node and Hopf bifurcation boundaries were determined from the slow-flow Jacobian, its eigenvalues, and the associated Routh–Hurwitz conditions. Post-Hopf periodic orbits were computed by time integration, and their stability was evaluated using Floquet multipliers obtained from the corresponding variational equations. All calculations were performed in MATLAB; further details are provided in **Supplementary Information 1 Section 1**.

## Acknowledgement

This work was supported by the Defense Advanced Research Projects Agency (DARPA) under Grants No. HR00112590104, HR00112590103, and HR00112390018. The views, opinions and/or findings expressed are those of the authors and should not be interpreted as representing the official views or policies of DARPA or the U.S. Government. The authors gratefully acknowledge the U.S. Army Research Laboratory for providing the reference clocks used in this study. The authors also thank Dr. David A. Howe for valuable technical discussions on frequency-stability metrology and time-deviation analysis.

## Author Contribution

J.G. and S.M. contributed equally to this work. J.G., S.M. and R.T. designed the devices. J.G. fabricated the devices. J.G., S.M. and S.M.E.H.Y. performed the experimental characterization. Y.K. and H.C. developed the theoretical model and performed the numerical simulations. J.G., S.M., Y.K., H.C. and R.T. analysed the data and wrote the manuscript. R.T. and H.C. supervised the work. All authors reviewed and approved the manuscript.

## Competing interests

The authors declare that they have no competing interests.

## Data availability

The data that support the findings of this study are available from the corresponding author upon reasonable request.

## Code Availability

The MATLAB scripts used for the slow-flow calculations, Jacobian and Routh–Hurwitz analysis, Floquet-multiplier calculations, and modified Allan-deviation processing are available from the corresponding author upon reasonable request.

## Corresponding author

The corresponding author is Roozbeh Tabrizian, *rtabrizi@umich.edu*.